\documentstyle[twocolumn,aps,prl,epsf,amssymb]{revtex}
\begin{document}
%
\draft

\wideabs{
\title{Elastic Compton scattering from the deuteron and nucleon 
polarizabilities}
\author{D.L.~Hornidge, B.J.~Warkentin, R.~Igarashi, J.C.~Bergstrom, 
E.L.~Hallin, N.R.~Kolb, R.E.~Pywell, D.M.~Skopik, J.M.~Vogt}
\address{Saskatchewan Accelerator Laboratory,
University of Saskatchewan,
Saskatoon, Saskatchewan, Canada, S7N 5C6}
\author{G.~Feldman}
\address{Department of Physics, The George Washington University, Washington, 
D.C. 20052}
\date{\today}
\maketitle
%
\begin{abstract}
Cross sections for elastic Compton scattering from the deuteron were measured
over the laboratory angles $\theta_\gamma=$ {35$^\circ$}--{150$^\circ$}.
Tagged photons in the laboratory energy range $E_{\gamma}=$ 84--105~MeV were 
scattered from liquid deuterium and detected
in the large-volume Boston University NaI (BUNI) spectrometer.  
Using the calculations of Levchuk and L'vov, along with the measured 
differential cross sections, the isospin-averaged
nucleon polarizabilities in the deuteron were estimated.  A best-fit
value of $(\overline{\alpha}-\overline{\beta})$=2.6$\pm$1.8 was 
determined, constrained by dispersion sum rules.  This is markedly 
different from the accepted value for the proton of 
$(\overline{\alpha}-\overline{\beta})_p$=10.0$\pm$1.5$\pm$0.9.
\end{abstract}


}

Elastic photon scattering from deuterium can, in principle, yield
basic information on the substructure of the deuteron and hence the
nucleons themselves.  Compton scattering from the proton has
been used extensively to determine the polarizabilities of
the proton (see Ref.~\cite{Mac-95} and references contained therein).
The electric 
($\overline{\alpha}_p$) and magnetic
($\overline{\beta}_p$) polarizabilities constitute the first-order
responses of the internal structure of the proton to externally
applied electric and magnetic fields.  

The current status of the proton polarizabilities
has been reported in Ref.~\cite{Mac-95}.
\begin{eqnarray}
	(\overline{\alpha}-\overline{\beta})_p & = & 10.0\pm1.5\pm0.9,
	\label{prot-diff}\\
	(\overline{\alpha}+\overline{\beta})_p & = & 15.2\pm2.6\pm0.2,
	\label{expt-sum}
\end{eqnarray}
where the first error is the combined statistical and systematic,
and the second is due to the model dependence of the 
dispersion-relation extraction method.  
The units of $\overline\alpha$ and $\overline\beta$ are $10^{-4}$~fm$^3$.  
There is also a dispersion sum
rule which relates the sum of the polarizabilities to the
nucleon photoabsorption cross section.  
The generally accepted result for the Baldin sum rule 
\cite{Pet-81} is
\begin{equation}
	(\overline{\alpha}+\overline{\beta})_p= 14.2 \pm 0.5, 
	\label{prot-sum}
\end{equation}
although recent reevaluations yield $13.69 \pm 0.14$ \cite{Bab-98}
and $14.0\pm (0.3-0.5)$ \cite{levchuk1}. 
Note that the experimental value (Eq.~\ref{expt-sum}) is in agreement with
the sum rule.  The polarizabilities, obtained from Eqs.~\ref{prot-diff} and
\ref{prot-sum}, are
\begin{eqnarray}
	\overline{\alpha}_p & = & 12.1\pm 0.8\pm 0.5,   \\
	\overline{\beta}_p  & = &  2.1\mp 0.8\mp 0.5.\nonumber
\end{eqnarray}

The status of the neutron polarizabilities is much less satisfactory.
The majority of measurements of the electric polarizability of the 
neutron have been done by low-energy neutron scattering from the 
Coulomb field of a heavy nucleus.  
There is considerable disagreement between the two most
recent measurements \cite{Sch-91,Koe-95}.
Schmiedmayer {\it et al.} \cite{Sch-91} reported a value for the
{\it static} electric polarizability of
$\alpha_n=12.0\pm1.5\pm2.0$, where the first uncertainty is
statistical and the second systematic.  
The difference between the static ($\alpha$) and the 
Compton ($\overline\alpha$) polarizability
is small for the neutron.
These data have been 
reinterpreted by Enik {\it et al.} \cite{Enik-97}, and they
have suggested that a value
of $\alpha_n=7-19$ was more appropriate.  In a separate experiment,
Koester {\it et al.} \cite{Koe-95} have reported a value of
$\alpha_n=0\pm5$.  Clearly, the current experimental value of 
$\alpha_n$ has large uncertainties.
Once $\alpha_n$ is obtained, $\beta_n$ can be determined via
the sum-rule relation for the neutron \cite{Pet-81},
\begin{equation}
	(\overline{\alpha}+\overline{\beta})_n = 15.8 \pm 0.5.
	\label{neut-sum}
\end{equation}
The recent reevaluations of this sum rule yield
$14.40 \pm 0.66$\cite{Bab-98}
and $15.2\pm (0.3-0.5)$ \cite{levchuk1}. 

An alternate method of measuring the neutron polarizabilities is 
through the use of the quasi-free Compton scattering reaction
$d(\gamma,\gamma^\prime\,n)p$ in which the scattered photon is 
detected in coincidence with the recoil neutron.
In certain kinematic regions, the proton behaves as a spectator 
and the scattering is primarily from the neutron.  There has been one
measurement reported on this reaction using bremsstrahlung photons with 
an endpoint of 130 MeV~\cite{Rose-90}.  However, due to poor
statistics, the resulting determination of $\overline{\alpha}_n$
effectively gives only an upper limit.
Experiments have recently been completed at Mainz \cite{Wiss-99}, LEGS,
and SAL, but results have yet to be reported.

A third method to determine the polarizability of the neutron is 
through the elastic Compton scattering reaction $d(\gamma,\gamma)d$.  
The only reported measurement of this reaction was
conducted at Illinois~\cite{Lucas-94}.  The low photon energy
of this experiment resulted in reduced sensitivity to the polarizabilities
and hence large error bars.
Since the deuteron amplitude is sensitive to the sum of
the proton and neutron polarizabilities, 
to obtain specific information on the neutron,
it is necessary to subtract the proton polarizabilities. 
More serious concerns are the
contributions from meson exchange currents and other nuclear effects.
Therefore, the final results for the neutron polarizabilities depend
on a model calculation as well as knowledge of the proton polarizabilities.

The present measurement was performed at the Sask\-atch\-ewan
Accelerator Laboratory 
(SAL).  The facility houses a 300~MeV linear accelerator (LINAC) that injects 
electrons into a pulse-stretcher 
ring (PSR), producing a nearly continuous wave (CW) electron beam.  The LINAC 
and PSR were used in conjunction with a high-resolution, high-rate photon 
tagger, a cryogenic target system, and a large-volume NaI detector.
Complete details of the experiment can be found in Ref.~\cite{horn99}.

The $d(\gamma,\gamma)d$ cross section was measured using
tagged photons in the energy range 84--105~MeV with a
resolution of 0.3--0.4~MeV.  
An electron beam of 135~MeV
and {$\sim$65\%} duty factor was incident on a 115~$\mu$m aluminum radiator
producing bremsstrahlung photons, which were
tagged via the standard photon tagging technique using
the SAL photon tagger \cite{vogt93}.
The average tagged flux was $\sim{6}\times{10^{7}}$ photons/s 
integrated over the photon energy range.
The tagging efficiency was measured approximately every 8~hours 
during the experiment by using
a lead glass $\check{\rm C}$erenkov detector directly in the beam 
to detect photons in coincidence with electrons in the focal plane.  
The tagging efficiency was approximately 53\%.

Photons scattered from the 12.7~cm long, liquid-deuterium target were 
detected in the large-volume Boston University NaI (BUNI) gamma-ray 
spectrometer.
BUNI is composed of five optically-isolated segments of NaI,
each 55.9~cm in length: the core (26.7~cm in diameter) and four
quadrants that form an 11.4~cm thick annulus around the core.
Since the inelastic contribution to $d(\gamma,\gamma^\prime)d$
begins only 2.2~MeV below the elastic peak, it was essential that
the photon detector have at least 2\% resolution at 100~MeV.
The excellent resolution of BUNI is mainly due to the fact that it
effectively contains 100\% of the 
electromagnetic showers created by the incident photons.
Scattered photons were detected at lab angles of 
$35^\circ$, $60^\circ$, $90^\circ$, $120^\circ$, and $150^\circ$.

A zero-degree (or {\em in-beam}) calibration of the detector
was done once, in the middle of 
the experiment, in order to obtain both the lineshape
of BUNI and an energy calibration for the NaI core.
The NaI quadrants were
calibrated daily with a radioactive source (Th-C).

\begin{figure}[htb]
\begin{center}
\epsfxsize=3.38in
\epsffile{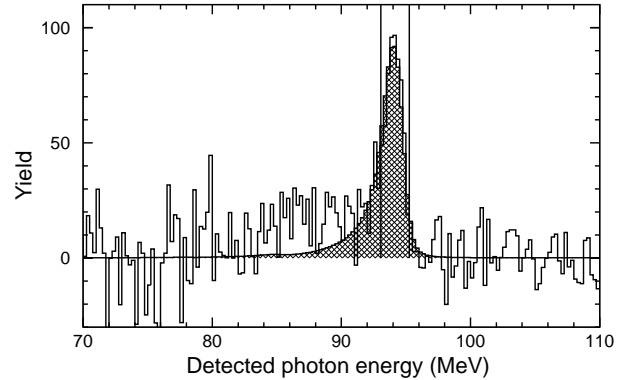}
\caption{Photon energy spectrum for $\theta_\gamma$= 150$^\circ$ 
showing the lineshape fit (shaded area) to the elastic peak.
The inelastic contribution is readily visible.  
The region of interest
is indicated by the vertical lines.}
\label{comparison}
\end{center}
\end{figure}

Figure~\ref{comparison} depicts the BUNI energy spectrum, after
randoms have been subtracted, summed over
the incident photon energy range.  The contribution from empty-target
backgrounds has also been subtracted.
Each channel in the focal plane of the tagger corresponds to
a different incident photon energy and hence a different energy
for the Compton scattered photon.  To sum over tagger channels, the
detected photon energy in BUNI was shifted to a value corresponding to
the maximum incident photon energy of 105~MeV.  The magnitude of the 
shift was determined by which tagger channel registered the photon.

After background subtraction,
the elastic peak in the BUNI energy spectrum was integrated 
over a specific region of interest (ROI),
depending on the angle, to obtain the yields.
This region was chosen to ensure that no inelastic contributions 
were contained in the integrated region.  
The ROI for the 150$^\circ$ photon energy spectrum is shown in 
Fig.~\ref{comparison} (vertical lines).  The inelastic contribution is
evident in the energy range 80--90~MeV.

Systematic errors associated with the energy calibration were
investigated by comparing the chi-square between the data and
the in-beam lineshape for different calibrations.  At the
95\% confidence level, the uncertainties in the cross sections
were 1--4\%, depending on angle.

The detection efficiency for the scattered photons 
was largely constrained by the small size of the ROI and 
photon absorption before reaching the detector.
Since the ROI (Fig.~\ref{comparison}) only selected the peak of the
elastic scattering distribution, a significant number of events from
the tail of the lineshape were excluded.
The ROI efficiency was deduced using the zero-degree
Compton spectrum, shifted to the appropriate energy, and
represents the response of monochromatic photons in BUNI.
Comparisons of measurements and EGS simulations of zero degree
and scattering geometries \cite{Warkentin-99} confirmed that 
the lineshape of BUNI measured at zero degrees was consistent 
with the measured lineshape in the scattering geometry within
$\sim$3\%.  The ratio of counts inside the ROI to the total counts 
in the lineshape yielded an efficiency of (62$\pm$2)\%.

The absorption efficiency was broken into two parts:  that
due to absorption of photons in the target and associated apparatus,
and that due to absorption of photons 
in the materials
located between the NaI crystal and the target enclosure.
The first absorption factor was obtained
with an EGS simulation \cite{Warkentin-99}.
The second was found by integrating
the entire zero-degree lineshape from the in-beam calibration
and comparing it to the incident
photon flux as determined by the photon tagger.
The absorption factor was approximately (89$\pm$1)\%, which gave
an overall detection efficiency of (55$\pm$2)\%, relatively
independent ($<$1\%) of scattering angle.

The effect of the high-rate photon flux on the measured
yield was investigated.  Rate effects were found to be 
$<$2\% in all cases \cite{horn99}.

Sources of systematic errors in this experiment included 
target thickness (2.5\%),
solid angle (1.6\%), detection efficiency (3.6\%), incident photon flux (1\%), 
and energy calibration (1--4\% depending on angle).  Adding
contributions in quadrature gives a total systematic error
of 6.4\% (35$^\circ$), 4.9\% (60$^\circ$), 4.8\% (90$^\circ$), 
4.9\% (120$^\circ$), and 5.2\% (150$^\circ$).

\begin{figure}[htb]
\begin{center}
\epsfxsize=3.38in
\epsffile{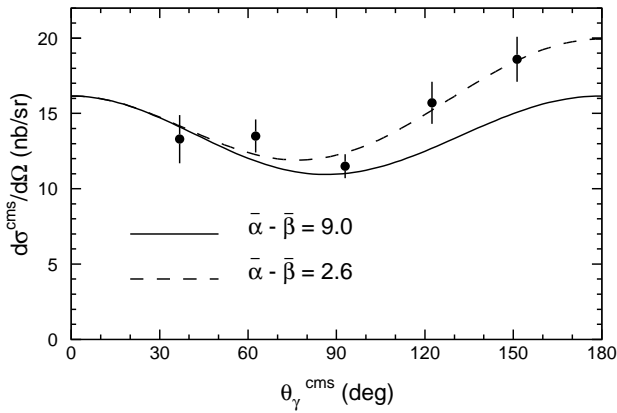}
\caption{Differential cross sections for elastic 
scattering of 84-105~MeV photons from deuterium. 
The curves are calculations of Levchuk and L'vov
with different polarizabilities.
}
\label{diffxs}
\end{center}
\end{figure}

The final center-of-mass system (CMS) differential cross sections
for the $d(\gamma,\gamma)d$ reaction as a function of CMS
scattering angle are displayed in Fig.~\ref{diffxs} and are listed
in Table~\ref{cross}.  The data were averaged over the incident
photon energy range of 84-105~MeV.  The error bars in the figure are the 
quadratic sum of the statistical and systematic errors.

\begin{figure}[htb]
\begin{center}
\epsfxsize=3.38in
\epsffile{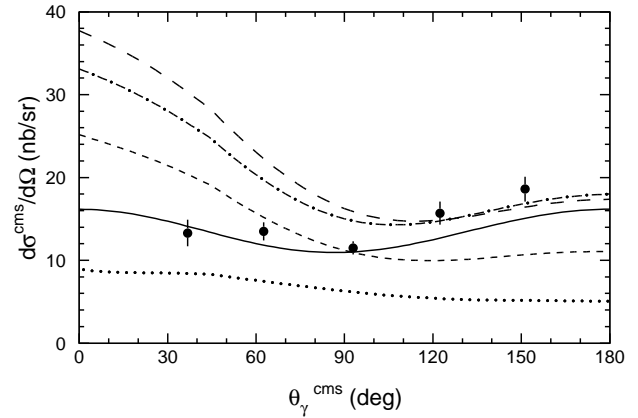}
\caption{Theoretical calculations of Levchuk and L'vov
for the $d(\gamma,\gamma)d$
differential cross section at E$_\gamma$=94~MeV.
The contributions, added successively, are
the resonant (dotted),
one-body seagull without polarizabilities (short-dashed),
two-body seagull (long-dashed),
resonant rescattering (dot-dashed), and
polarizabilities (solid).}
\label{ingredients}
\end{center}
\end{figure}

Levchuk and L'vov \cite{levchuk1,levchuk2} have calculated the 
differential cross section
for the $d(\gamma,\gamma)d$ reaction in the framework of a diagrammatic
approach.  The scattering amplitude is expressed in terms of resonance
and seagull amplitudes.  The resonance amplitudes correspond to two-step 
processes and include rescattering of the intermediate nucleons.  
The one- and two-body seagull amplitudes 
involve a photon being absorbed and emitted at the
same moment within the energy scale involved.  
The one-body seagull diagrams include the nucleon polarizabilities.
The contribution of the various ingredients to the $d(\gamma,\gamma)d$
reaction at 94~MeV is illustrated in Fig.~\ref{ingredients}.

The solid line in Figs.~\ref{diffxs} and \ref{ingredients} is the full 
calculation of
Levchuk and L'vov with nominal free values of the isospin-averaged
polarizabilities, $\overline{\alpha}=12$.0, $\overline{\beta}=3.0$ or 
$(\overline{\alpha}-\overline{\beta})=9.0$.  
By imposing the sum-rule constraints (Eqs.~\ref{prot-sum} 
and \ref{neut-sum}), the difference of the
polarizabilities can be taken as a free parameter and fitted to the
experimental data.  
The dashed curve in Fig.~\ref{diffxs} corresponds to the best-fit value of 
\begin{equation}
	(\overline{\alpha}-\overline{\beta})=2.6\pm1.8,
	\label{neut-diff}
\end{equation}
which was determined by 
minimizing the chi-square ($\chi^2/N_{d.o.f.}=2.8/4$)
between the calculation and the data.
Choosing the sum-rule results of Babusci {\it et al.}
gives a $<$4\% increase in 
$(\overline{\alpha}-\overline{\beta})$.
The fitted value is substantially
different from the nominal free value,
and is driven by the back-angle cross sections.  
The extracted $(\overline{\alpha}-\overline{\beta})$ may be
interpreted in terms of (1)~medium modifications to the nucleon
polarizabilities ($\Delta\overline{\alpha}$ and $\Delta\overline{\beta}$),
(2)~neutron polarizabilities 
which are quite different from the proton, or
(3)~missing physics in the calculation used
to extract the polarizabilities.
It is important to note that the extracted value
of $(\overline{\alpha}-\overline{\beta})$ is
dependent upon the theoretical model used.
For a discussion of the model dependence of the
existing calculations see Refs.~\cite{levchuk1,levchuk2}.

Medium modifications to the free polarizabilities for
Compton scattering from light nuclei have been postulated 
\cite{Eric-87,Buna-92}.
Recent measurements on {$^4$He} \cite{Fuhr-95}, 
{$^{12}$C} \cite{Warkentin-99}, and {$^{16}$O} \cite{Feld-96} 
have suggested that changes on the order of
$\Delta\overline{\beta}$ = $-\Delta\overline{\alpha}$ =
4--8 were required in order to describe the data.  These differences
were driven by the back-angle cross sections which tended to be
larger than the theoretical predictions, as is also seen in the
current $d(\gamma,\gamma)d$ results.  Other
measurements, performed at Lund, have reported no
modification of the free polarizabilities \cite{lund}.
Clearly this issue has yet to be resolved.  In the current
measurement, modifications of $\Delta\overline{\beta}\sim$3
would be required to
explain the data.  However, for a lightly bound system like 
the deuteron, medium modifications might be expected to be small.

Assuming that the proton polarizabilities are unmodified in the
deuteron, the neutron polarizabilities can be extracted from the fitted 
$(\overline{\alpha}-\overline{\beta})$.
Using the fitted value along with Eq.~\ref{prot-diff}
yields 
\begin{equation}
	(\overline{\alpha}-\overline{\beta})_n = -4.8 \pm 3.9,
\end{equation}
which can be used along with Eq.~\ref{neut-sum} to determine
the neutron polarizabilities,
\begin{eqnarray}
	\overline{\alpha}_n & = &  5.5 \pm 2.0,
	\label{results} \\
	\overline{\beta}_n  & = & 10.3 \mp 2.0. \nonumber
\end{eqnarray}
The error bars do not include any model dependence introduced
by the theoretical calculation and are anti-correlated
due to application of the sum rule.
These results are surprising, since the neutron values are not expected
to be radically different from the proton.  

Evidence for modification of the backward spin polarizability 
($\gamma_\pi$) has been reported \cite{tonnison}.  The suggested 
modification would increase the theoretical back-angle cross 
sections (Fig.~\ref{diffxs}) and hence increase the extracted 
$(\overline{\alpha}-\overline{\beta})$ by $\sim$3.
However, a recent measurement of $d(\gamma,\gamma^\prime p)n$
reports no evidence for such a modification of $\gamma_\pi$ 
\cite{Wiss2}.

There has also been a recent calculation of the $d(\gamma,\gamma)d$
differential cross section within the framework of baryon chiral
perturbation theory \cite{Bea-99}.  The results of the calculation
of Beane {\it et al.},
which includes $O(Q^3)$ terms and only a partial set of the $O(Q^4)$
corrections, are similar in magnitude to that of Levchuk and L'vov.
However, Beane {\it et al}.\ warn that their prediction at 95~MeV
has considerable uncertainty due to the slow convergence of the
series.  Understanding what physics may be missing in the theoretical
calculations will have to wait for the full $O(Q^4)$ treatment.
\begin{table}
\caption{Center-of-mass differential cross sections for the elastic
scattering of 84--105~MeV photons from deuterium.
The first error is statistical and the second is systematic.}
\label{cross}
\begin{tabular}{ddc}
$\theta_{\gamma}^{lab}$ & $\theta_{\gamma}^{cms}$ & 
d$\sigma^{cms}$/d$\Omega_\gamma$ \\
(deg) & (deg) & (nb/sr) \\
35.0  & 36.8  &  13.3 $\pm$ 1.3 $\pm$ 0.9 \\
60.0  & 62.7  &  13.5 $\pm$ 0.8 $\pm$ 0.7 \\
90.0  & 93.0  &  11.5 $\pm$ 0.6 $\pm$ 0.6 \\
120.0 & 122.6 &  15.7 $\pm$ 1.1 $\pm$ 0.8 \\
150.0 & 151.5 &  18.6 $\pm$ 1.1 $\pm$ 1.0 \\
\end{tabular}
\end{table}

Other theoretical calculations for the $d(\gamma,\gamma)d$
reaction have been reported.  The calculations of Karakowski
and Miller \cite{Kara} underpredict the present measurement by 
a factor of $\sim$2 at backward angles when using the nominal free values
of the polarizabilities.
Wilbois {\it et al.} \cite{Wilb} have reported a calculation at 
100~MeV but with polarizabilities set to zero.  Finally,
Chen {\it et al.} \cite{Chen} have reported results only up to 
a photon energy of 69~MeV.

Forthcoming results on the quasi-free Compton scattering from 
the deuteron, with detection of the recoiling nucleon,
should shed some light on this controversy.  By doing the
experiments in quasi-free kinematics 
(in the energy region E$_\gamma$=200--300~MeV
and backward angles for the scattered photons)
the model dependence is minimized, and
the polarizabilities can be
extracted separately for the proton and the neutron.

The authors would like to thank M.I.~Levchuk  and A.I.~L'vov
for supplying the code for their theoretical calculations.

This work was supported in part by a grant from the Natural Science 
and Engineering Research Council of Canada.

%


\end{document}